\documentclass[final,5p,twocolumn]{elsarticle}
\usepackage{subfigure}
\usepackage{gensymb}
\usepackage[T1]{fontenc}
\usepackage{multicol}
\usepackage{textcomp}
\usepackage[utf8]{inputenc}
\usepackage[ngerman, english]{babel}
\usepackage{amsmath}
\usepackage{amssymb}
\usepackage{graphicx}
\usepackage{float}
\usepackage[breaklinks=true]{hyperref}
\usepackage{siunitx}
\usepackage{caption}
\usepackage{color}
\usepackage{bm}

\usepackage{url}
\usepackage{comment}

\usepackage{listings}
\usepackage{xcolor}
\definecolor{codegreen}{rgb}{0,0.54,0}
\definecolor{codegray}{rgb}{0.54,0.0,0.0}
\definecolor{codepurple}{rgb}{0.54,0.0,0.1}
\definecolor{backcolour}{rgb}{0.95,0.95,0.92}
\lstdefinestyle{mystyle}{
    commentstyle=\color{codegreen},
    keywordstyle=\color{magenta},
    numberstyle=\tiny\color{codegray},
    stringstyle=\color{codepurple},
    basicstyle=\ttfamily\footnotesize,
    breakatwhitespace=false,         
    breaklines=true,                 
    captionpos=b,                    
    keepspaces=true,                 
    numbersep=5pt,                  
    showspaces=false,                
    showstringspaces=false,
    showtabs=false,                  
    tabsize=2
}
\usepackage{lineno}
\usepackage{mathtools, cuted}
\hypersetup{draft}

\journal{Computer Physics Communications}

\begin{document}

\begin{frontmatter} 

\title{NTMpy: An open source package for solving coupled parabolic differential equations in the framework of the three-temperature model}

\author[1]{Lukas Alber}
\author[2]{Valentino Scalera}
\author[1]{Vivek Unikandanunni}
\author[3]{Daniel Schick}
\author[1,4]{Stefano Bonetti}

\address[1]{Department of Physics, Stockholm University, 106 91 Stockholm, Sweden}
\address[2]{Department of Electrical Engineering and ICT, University of Naples Federico II, 80125 Naples, Italy.}
\address[3]{Max Born Institute for Nonlinear Optics and Short Pulse Spectroscopy, Max-Born-Straße 2A, 12489 Berlin, Germany}
\address[4]{Department of Molecular Sciences and Nanosystems, Ca' Foscari University of Venice, 30172 Venezia-Mestre, Italy}


\begin{abstract}
The NTMpy code package allows for simulating the one-dimensional thermal response of multilayer samples after optical excitation, as in a typical pump-probe experiment. Several Python routines are combined and optimized to solve coupled heat diffusion equations in one dimension, on arbitrary piecewise homogeneous material stacks, in the framework of the so-called three-temperature model. The energy source deposited in the material is modelled as a light pulse of arbitrary cross-section and temporal profile. 
A transfer matrix method enables the calculation of realistic light absorption in presence of scattering interfaces as in multilayer samples. The open source code is fully object-oriented to enable a user-friendly and intuitive interface for adjusting the physically relevant input parameters. Here, we describe the mathematical background of the code, we lay out the workflow, and we validate the functionality of our package by comparing it to commercial software, as well as to experimental transient reflectivity data recorded in a pump-probe experiment with femtosecond light pulses.
\end{abstract}


\end{frontmatter}

{\bf Program summary}

\begin{small}
\noindent
{\em Program title}: NTMpy v.0.1.1       \\
{\em Program obtainable from}: Github.com \url{https://github.com/udcm-su/NTMpy} \\
{\em Licensing provisions}: MIT license (MIT) \\
{\em No. of lines in distributed program, including test data, etc.}: 1804 \\
{\em No. of bytes in distributed program, including test data, etc.}: 106212 \\
{\em Distribution format}:(tar.gz) \\ 
{\em Programming language}: Python \\
{\em Computer}: PC, Workstation \\
{\em Operating system}: Running Python installation required. (Operating system independent) \\
{\em RAM}: Typically in the range of hundrets of megabytes \\
{\em External routines}: Python 3.5 or higher, numpy, matplotlib, bsplines, tqdm \\
{\em Nature of problem}: 1-dimensional coupled non linear partial differential equations; diffusion and relaxation dynamics formultiple systems and multiple layers. \\
{\em Solution method}: Simulate the diffusion and relaxation dynamics of up to 3 coupled systems via an object oriented user interface. In order to approximate the solution and its derivatives in space B-Spline interpolation is used. The solution is developed in time via the Explicit Euler method. \\
{\em Unusual features}: A routine to automatically select the ideal time step for stability of the algorithm is implemented. Routines for output of raw data in order to post process and pre- made visualization routines are implemented. \\
{\em Running time}: Main influence factor is how many time steps are required to integrate the equation in time. The time step evaluation in the Explicit Euler loop depends on the input parameters of the equation. Almost all tested problems can be solved under 30 seconds on a recent computer.
\end{small}

\section{Introduction}\label{sec:Introduction}
The understanding of how heat is transported at the nanometer level and at ultrafast time scales in different materials is an open question in modern condensed matter research. The increasing availability of commercial femtosecond laser systems makes it possible to study material dynamics at sub-picosecond time scales, allowing for investigating non-equilibrium energy transport. This also asks for numerical computations able to model the experimental evidence and to either validate or extract a set of physical parameters.

A commonly used strategy to describe the highly non-equilibrium processes induced by ultrafast laser excitation, is the $N$-temperature model ($N$ = 2 or 3, typically) posed by Anisimov et al. \cite{Anisimov}. This model is a set of coupled parabolic differential equations, which describe the temporal evolution of the energy of the electron, lattice, and spin systems as well as the transfer of energy between these systems. The $N$-temperature model is used to simulate a broad range of experiments in the ultrafast community, e.g.  Ref. \cite{Bigot,KiethNelsonTG,koopmans2010explaining,Schick2014}, but an open source implementation of it is still missing. 

Here, we provide an open source, user-friendly Python package able to solve the $N$-temperature model in one-dimension, with arbitrary, piece-wise homogeneous layers with different physical properties \cite{NTMpy}. The time-step selection to optimize the solving routines and guarantee their convergence, the calculation of the deposited energy by an arbitrarily shaped laser pulse using the transfer matrix method, and visualization routines, are all automatized for the early-user. However, the object oriented design of the package, also allows for easy customization for the advanced users.

After laying out the mathematical background on which the implementation of the solver is based on in Sec. \ref{sec:MathMethods}, we introduce the workflow of the program by showing the most important commands and the output, users can expect in Sec. ~\ref{sec: Implementation}. To validate our results we compare the computation of this open source package to commercial software, and further more to real experimental data, in Sec. ~\ref{sec:Example}.

\section{Mathematical Methods and Background}\label{sec:MathMethods}

When a material is illuminated by light, the energy of the electromagnetic radiation is partly reflected, partly transmitted through the material, and partly absorbed by it. The absorbed electromagnetic energy essentially heats the material, rising its temperature. While this is a simple problem to solve for the case of a continuous light source, it becomes immediately more complex for the case of an ultrashort laser pulses, i.e. with sub-picosecond pulse duration. In this case, the concept of temperature is in itself an ill-defined one, because the different energy reservoirs in a material (the electronic $E$ and lattice $L$ systems, and also the spin $S$ system for a magnetically ordered sample) responds on different time scales and are not in equilibrium among each other. Generally, only the electrons can react fast enough to the ultrashort laser pulse, and the energy is released from the electronic system to the other heat reservoirs only at a later time.

To treat this problem, three reservoirs are considered, in order to allow for the definition of a temperature and to solve the heat equation in time, and then coupled to allow for the energy to flow between them. 
This is addressed by considering three independent reservoirs, in order to allow for the definition of a temperature and to solve the heat equation in time in each of the systems. Finally, the three systems are coupled such that the energy can flow between them. This idea is generalized to a $N$-temperature model, describing the heat exchange between the systems and the heat diffusion along the one-dimensional, multiple layered material. 
Each of the systems - electron, lattice, and spin - has its individual temperature $T^{E,L,S}(x,t)$, where $x$ denotes the depth in the specimen and $t$ is the time.

The dynamics of every system are described by a parabolic partial differential equation, namely

\begin{align}\label{eq: N-temperature model}
    \begin{cases}
        C^E(T^E)\cdot\rho\cdot\partial_tT^E = \partial_x\left(k^E(T^E)\cdot \partial_xT^E\right) + \\
        \qquad G^{EL}\cdot(T^L-T^E)+G^{SE}\cdot(T^S-T^E) + S(x,t)\, , \\
        C^L(T^L)\cdot\rho\cdot\partial_tT^L = \partial_x\left(k^L(T^L)\cdot \partial_xT^L\right) + \\
        \qquad G^{EL}\cdot(T^E-T^L)+G^{LS}\cdot(T^S-T^L)\, , \\ 
        C^S(T^S)\cdot\rho\cdot\partial_tT^S = \partial_x\left(k^S(T^S)\cdot \partial_xT^S\right) + \\
        \qquad G^{SE}\cdot(T^E-T^S)+G^{LS}\cdot(T^L-T^S)\, .
    \end{cases}
\end{align}

Here $C^{E,L,S}(T)$ and $k^{E,L,S}(T)$ are the specific heat capacity and thermal conductivity, which can be defined as constants or as functions of the respective temperature $T^{E,L,S}$. Since we consider a stack of multiple layers $C^i$ and $k^i$ also depend on the respective layer, hence they are $C^i_s(T^i)$ and $k^i_s(T^i)$, the dependence on the layer $s$ will be omitted when not necessary in order to have a simpler notation.\\ The term $S(x,t)$ is responsible for the heat injection to the electronic system and physically corresponds to a pulsed laser source hitting the sample at the surface. Note that in Eq. \eqref{eq: N-temperature model}, we assume the coupling $G$, responsible for the heat exchange and relaxation between the systems, to be linear as in the formulation from Anisimov \textit{et al.}, Ref. \cite{HohlfeldTTM}, \cite{NorrisFemptosecond} and other groups. However, it should be mentioned, that considering a non-constant $G(T)$ is current subject to research \cite{DFT}.

The temperature profile is expressed as a linear combination of basis functions $B_m(x)$ for $m=1,\ldots, M$  with time depending coefficients for every subsystem, $i\in \lbrace E,L,S \rbrace$, $c^i_{m}(t)$, i.e.
\begin{equation}\label{eq: BSpline decomposition}
T^i(x,t) = \sum_{m=1}^M c^i_{m}(t) B_m(x)\ .
\end{equation}
The solution of Eq.~\eqref{eq: N-temperature model} is  now reduced to a finite dimensional problem and consequently the diffusion equation cannot generally be solved exactly in the whole domain.

The solution can be approximated by imposing Eq.~\eqref{eq: N-temperature model} to be satisfied on a given grid of points $\lbrace x_1, x_2, x_3,\ldots, x_{M-1}\rbrace$, i.e. \textit{collocation points}. Two additional points $x_0$ and $x_M$ are  added to the grid on the boundary of the domain to impose the boundary conditions.

The temperatures and their derivatives have an exact analytic expression at the grid points
\begin{equation}\label{eq: differentiation matrices}
\begin{aligned}
T^i(x_j,t) &=\sum_{m=1}^M c^i_{m}(t)B_m(x_j)\, , \\ 
\frac{\partial T^i}{\partial x}(x_j,t) &=\sum_{m=1}^M c^i_{m}(t)\frac{\partial B_m}{\partial x}(x_j)\, ,\\ 
\frac{\partial^2 T^i}{\partial x^2}(x_j,t) &=\sum_{m=1}^M c^i_{m}(t)\frac{\partial^2 B_m}{\partial x^2}(x_j)\, .
\end{aligned}
\end{equation}
By introducing the $M\times M$ matrices $D_0$, $D_1$ and $D_2$ with generic elements
\begin{equation}
\begin{aligned}
\lbrace D_0\rbrace_{jm} &=B_m(x_j)\, ,\\ 
\lbrace D_1\rbrace_{jm} &=\frac{\partial B_m}{\partial x}(x_j)\, , \\ 
\lbrace D_2\rbrace_{jm} &=\frac{\partial^2 B_m}{\partial x^2}(x_j)\, ,
\end{aligned}
\end{equation}
the temperatures and their derivatives can be obtained by matrix products.
By using the Leibnitz formula Eq.~\eqref{eq: N-temperature model} can be the reformulated as follows
\begin{multline}\label{eq: finite elements}
\rho C^iD_0\frac{\mathrm{d} }{\mathrm{d} t}\bm c^i= \\  
k^i(D_0\bm c^i)\cdot D_2\bm c^i+\frac{\mathrm{d}k_s^i}{\mathrm{d}(D_0\bm c^i)}\cdot(D_1\bm c^i)^2+S^i(x,t) + \\ 
\sum\limits_{\substack{k\in \lbrace E,L,S\rbrace\\k\neq i}}\ G^{ik}(D_0(\bm c^k -\bm c^i))\, ,
\end{multline}
for $i=\lbrace E,L,S\rbrace$, such that $\bm c_i = \lbrace c^i_{1},c^i_{2},\ldots, c^i_{M}\rbrace$ is the vector of coefficients relative to the $i$-th temperature, the dot ($\cdot$) denotes the elementwise product of two vectors and $(\bm a)^2$ denotes the vector whose elements are the squares of the elements of the vector $\bm a$.

The time evolution of the coefficient vectors $\bm c^i$ is evaluated using the explicit Euler formula, which reads 

\begin{equation}\label{eq: explicit Euler}
D_0\bm c^i(t+\Delta t) = D_0\bm c^i(t) + \Delta t D_0 \frac{\mathrm{d} \bm c^i}{\mathrm{d} t}(t)\, ,
\end{equation}
where the time derivative is calculated as shown in Eq.~\eqref{eq: finite elements} and $S$ is present only for $i=E$.

When an analytical formula for $\mathrm{d}k^i/\mathrm{d}T^i$ is unavailable, this derivative  can be computed numerically without introducing significant errors since the conductivity is typically a regular function.
Equation~\eqref{eq: explicit Euler} must be completed with the initial- and boundary conditions at the left and right end of the material under consideration.

The boundary conditions in this software can be of Dirichlet type or of a modified Neumann type according to whether the temperature or the heat flux is assigned at the boundaries. Hence we have the two following options for the boundary conditions

\begin{multline*}
\text{Dirichlet}:\begin{cases}
(D_0)[\ 0,:]\ \bm c^i = T_{BC}(x_0,t)
\\
(D_0)[M,:]\ \bm c^i = T_{BC}(x_M,t)
\end{cases}\, , \text{ or }\\
\text{Neumann}:\begin{cases}
k^i(T^i(x_0\,))((D_0\bm c^i)D_1)[\ 0,:]\ \bm c^i = H_{BC}(x_0,t)
\\
k^i(T^i(x_M))((D_0\bm c^i)D_1)[M,:]\ \bm c^i = H_{BC}(x_M,t)\, ,
\end{cases}
\end{multline*}

where the notation $[\ 0,:]$ and $[M,:]$ indicates the first and the last row of the matrix and $T_{BC}$ or $H_{BC}$ are the assigned boundary conditions.

\begin{figure}[b]
    \centering
    \includegraphics[scale = 0.22]{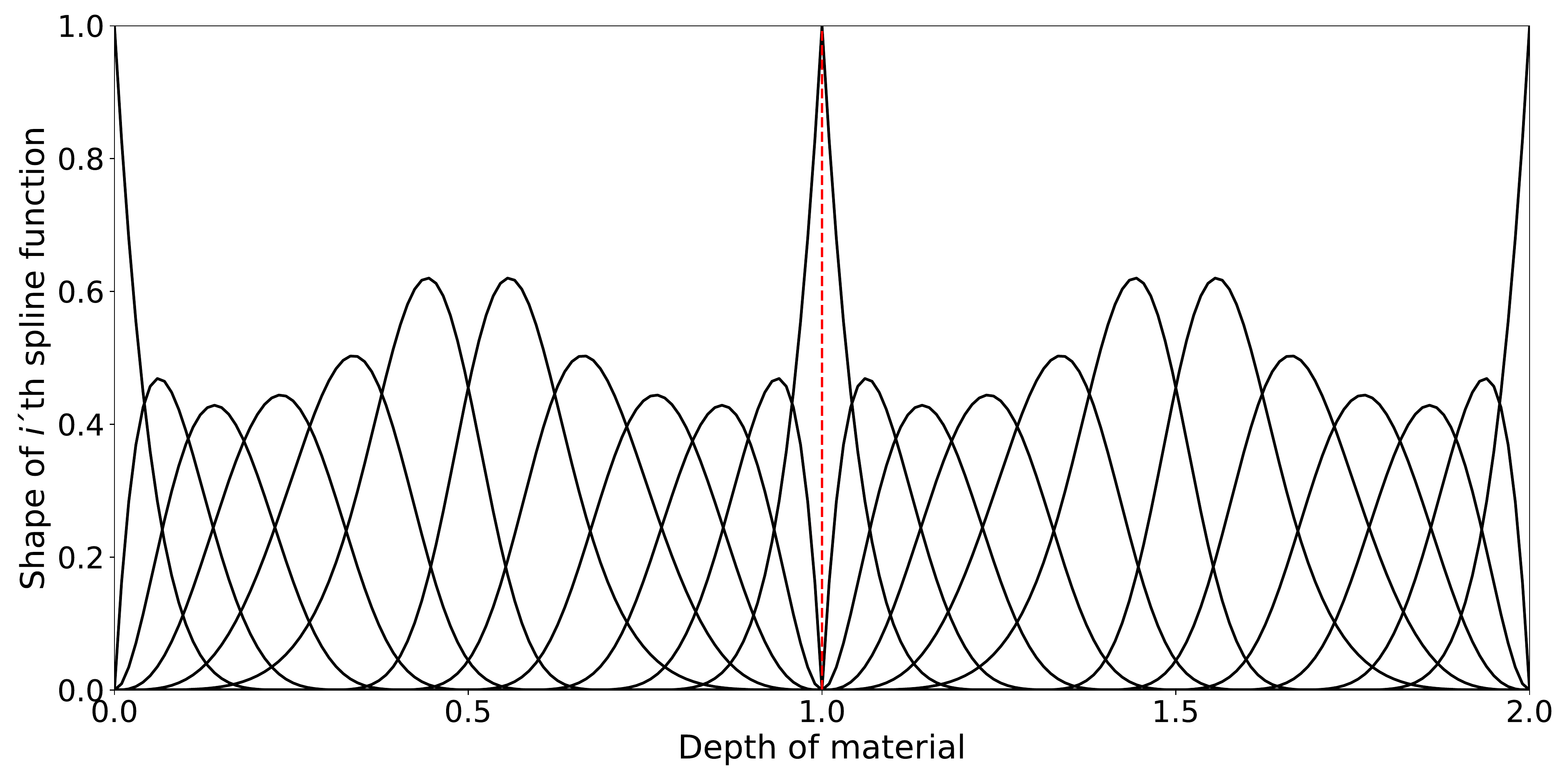}
    \caption{A set of order 5-continuous B-Splines for a two layer system, used to represent the solution in space, where each layer has a length of 1.}
    \label{fig: BSplines}
\end{figure}

The basis functions chosen for the approximation of the solution are B-Splines realized with the Cox-de Boor algorithm. 
These splines are continuous and have continuous derivatives up to a chosen order. A set of B-Spline basis functions, as used in the software, are depicted in Fig. \ref{fig: BSplines}, where a stack of two layers is under consideration.

The use of $\mathcal{C}^k$ -continuous functions is justified by the nature of the physical problem: the presence of a discontinuity in the temperature would  cause an infinite heat flux, while a discontinuity in its first derivative would imply a finite heat flux into an infinitesimal control volume unless the conductivity is discontinuous.

Discontinuities in the conductivity can be present when the specimen is made of two or more different materials stacked together. In this case the heat flowing into the interface between the $s$-th and the $(s+1)$-th layer, for the $i$-th temperature is
\begin{equation}\label{eq: interface heat flow}
H_\text{Interface}=
k_{i,s}(T^i)\lim_{x\rightarrow x_I^-}\partial_xT^i-k_{i,s+1}(T^i)\lim_{x\rightarrow x_I^+}\partial_xT^i=0\, ,
\end{equation}
where the second subscript of $k$ indicates the layer, $x_I$ is the position of the interface and the superscript $+$ and $-$ indicates the limit from right and the left respectively.

In order to correctly represent the temperatures at the interface a different set of B-Spline is considered for each layer (see Fig. \ref{fig: BSplines}), and the continuity of the temperature and the condition Eq. \eqref{eq: interface heat flow}, i.e. the conservation of the heat flow, are imposed at the interface.

Notice that when the coefficients $c_{im}$ are determined the limit and derivative appearing in Eq.~\eqref{eq: interface heat flow} are analytically computed.

\ifx
For the exchange term $G_i$ in equation \eqref{eq: N-temperature model}, we consider a linear function of the temperature differences at the same point
\begin{equation}
G_i=\sum_{\substack{\ell=1\\j\neq i}}^N G_{i\ell}(\varphi^\ell-\varphi^i) = \sum_{\substack{j=1\\j\neq i}}^N G_{ij}D_0(\bm c_j-\bm c_i)
\end{equation}
where $G_{i\ell}$ for $(i,\ell)= 1,\ldots,\N$ are constant terms. For $N=2$ or $N=3$ we have $G_{i\ell}=G_{\ell i}$ because of the energy conservation.
\fi

\subsection{Evaluation of the Time Step}\label{sec: time step}
A crucial point for the precision, speed and stability of the code is the choice of an appropriate time step. When this is not supplied by the user, the time step is automatically determined according to a criterion guaranteeing stability on one side, but keeping the running time of the simulation as small as possible on the other side.

For a linear $N$-temperature system with a single layer the dynamical matrix is given by 
\begin{equation}\label{eq: dynamic matrix}
\frac{1}{\rho}\text{diag}\left\lbrace \frac{k^E}{C^E}, \frac{k^L}{C^L}, \frac{k^S}{C^S}\right\rbrace\otimes (D_0^{-1}D_2)
+\frac{1}{\rho}\,M
\otimes I_N\, ,
\end{equation}
with
\begin{equation}
M = \left[\begin{array}{ccc}
- \dfrac{G^{EL}+G^{SE}}{C^E}& \dfrac{G^{EL}}{C^E}&\dfrac{G^{SE}}{C^E}\\[8pt]
\dfrac{G^{EL}}{C^L}&-\dfrac{ G^{EL}+G^{LS}}{C^L}&\dfrac{G^{LS}}{C^L}\\[8pt]
\dfrac{G^{SE}}{C^S}&\dfrac{G^{LS}}{C^S}&- \dfrac{G^{LS}+G^{SE}}{C^S}
\end{array}\right]
\end{equation}
where $I_N$ is the $N$ dimensional identity matrix and $\otimes$ denotes the Kronecker product.
In this system, to preserve stability, the time step needs to satisfy the condition
\begin{equation}\label{eq: stability condition}
\Delta t < \frac{2}{|\lambda|_{max}}\, ,
\end{equation}
where $|\lambda|_{max}$ is the eigenvalue of the dynamical matrix with the largest absolute value. We note that, $\Delta t$ depends on the input parameters $C$, $k$, $G$ and on the desired spatial resolution, represented in $D_0$ and $D_2$, but not on the heating source term $S(x,t)$.
 
In the nonlinear context we can still use condition \eqref{eq: stability condition} with the eigenvalues of an adapted version of matrix \eqref{eq: dynamic matrix}. We consider the worst scenario for stability  represented by contemporaneous large $k^i$'s (thermal conductivities) and small $C^i$'s (specific heats), which produce large eigenvalues in absolute value. To this purpose, in the worst scenario matrix, the $k^i$'s are replaced by their maximum, evaluated on a set of values of $T^i$ and the $C^i$'s are replaced by their minimum evaluated on the same set. This yields the following matrix
\begin{equation}\label{eq: dynamic matrix minmax}
\frac{1}{\rho}\text{diag}\left\lbrace \frac{k^E_{max}}{C^E_{min}\rho}, \frac{k_{max}^L}{C_{min}^L\rho}, \frac{k_{max}^S}{C_{min}^S\rho}\right\rbrace\otimes (D_0^{-1}D_2)
+\frac{1}{\rho}\,M_{max}
\otimes I_N\, ,
\end{equation}
where
\begin{equation}
M_{max} = 
\left[
\begin{array}{ccc}
- \dfrac{G^{EL}+G^{SE}}{C^E_{min}}& \dfrac{G^{EL}}{C^E_{min}}&\dfrac{G^{SE}}{C^E_{min}}\\[12pt]
\dfrac{G^{EL}}{C^L_{min}}&-\dfrac{ G^{EL}+G^{LS}}{C^L_{min}}&\dfrac{G^{LS}}{C^L_{min}}\\[12pt]
\dfrac{G^{SE}}{C^S_{min}}&\dfrac{G^{LS}}{C^S_{min}}&- \dfrac{G^{LS}+G^{SE}}{C^S_{min}}
\end{array}
\right]
\end{equation}
As this linearization, from $k_i(T)$ to $k_i^{max}$, depends on the set of values chosen for $T$, typical values of temperatures reached in the experiments on ultrafast dynamics are considered in the default case. That is, $T \in$ [270,3000]K, but they can be changed by the user via
{\ttfamily sim.stability\_lim([lowlimit,highlimit])}. Considering the worst scenario allows to evaluate the time step only once at the beginning of the simulation.

In multilayer systems the above procedure is repeated for each layer obtaining $\Delta t_1,\Delta t_2,\ldots, \Delta t_L$ where $L$ is the number of layers, the time step is $\Delta t = \min\lbrace\Delta t_1,\Delta t_2,\ldots, \Delta t_L\rbrace$. 

\section{Background and Implementation}
\label{sec: Implementation}
The code is implemented in Python with dependence on the \textit{numpy} and the \textit{bspline} package for the numerical computation, on the \textit{matplotlib} library for plotting of the results and the \textit{progressbar} package to monitor the elapsed time. Installation of the NTMpy package will automatically check if the dependencies are fulfilled and download the additional software if not.

In order to make the user interface friendly, the code is object oriented and it can be used either with command line or in a script. The package contains three main classes which are {\ttfamily source}, {\ttfamily simulation}, and {\ttfamily visual}: While the {\ttfamily source} class is a collection of methods for the generation of the energy injection matrix, the {\ttfamily simulation} class handles the computation of the solution and builds the core of the program. After a solution has been found, it can be passed on to the {\ttfamily visual} classes allowing a fast and easy depiction of the results respectively.

\subsection{Data Input}
Even though in principle the source function $S(x,t)$, injecting heat in space and time, can be of generic type, the most common types of sources are already defined in the code and one needs to specify only the main properties. That is, the user can choose between Lambert Beer's law or the transfer matrix method (TMM) to calculate the absorption profile in space and independently from that select either a Gaussian, a repeated Gaussian or even a custom time profile to evaluate the shape of the heating source in time.

For example a source with a Gaussian profile in time and exponential decay in space, considering multiple reflections, incident angle and polarization, i.e. TMM, is introduced with the following lines of code. 
\begin{lstlisting}[language=Python]
#Define a Source
s                   = source()
# Set source type
s.spaceprofile      = 'TMM'
s.timeprofile       = 'Gaussian' 
# Width of the Gaussian (in s)
s.FWHM              = 0.1e-12
# Area under the Gaussian (in J/m^2)
s.fluence           = 6*1e-3/1e-2**2 
# Set the time of the Gaussian peak (in s)
s.t0                = 1e-12 
# Wavelength in vacuum (in nm)
s.lambda_vac        = 400 
# Incident angle (in rad)
# (0 is perpendicular to the surface)
s.theta_in          = pi/4 
s.polarization      = 'p'
\end{lstlisting}

Note, that there is also a predefined way to calculate the spacial absorption according to Lambert Beer´s law via {\ttfamily s.spaceprofile = 'LB'}, which follows, except for the input of the incident angle and the polarization of the light, the same commands as above.
Alternatively a custom profile in time can be given for the source, if arrays of data, here {\ttfamily my\_time} and {\ttfamily my\_intensity}, are provided to the program. Now considering the Lambert Beer decay law in space, the commands to initialize a customized source are
\begin{lstlisting}[language=Python]
# Set source type
s.spaceprofile = 'LB'
s.timeprofile  = 'Custom'
# Set the value for the source
s.loadData = [my_time, my_intensity]
\end{lstlisting}

After the initialization of a source, one can proceed to initialize the simulation object, providing material specific parameters for every layer of the stack under consideration and then run the simulation.
The constructor of the simulation class has a mandatory input which is the number {\ttfamily N} of temperatures under consideration and an optional input which is the source {\ttfamily s}.
\begin{lstlisting}[language=Python]
# Define the simulation object
sim = simulation(N, s)
\end{lstlisting}
{\ttfamily N} can be 1, 2 or 3. When the properties of {\ttfamily s} are not specified as shown above, by default it is assumed that there is no source term in Eq. \eqref{eq: N-temperature model}, i.e. the fluence is 0.\\
The properties of the media are assigned by adding sequentially the layers of the materials with the method  {\ttfamily addLayer()}. For instance  
\begin{lstlisting}[language=Python]
# Add a layer  for a 2 temperature model
sim.addLayer(length, n, [k1, k2], 
                [C1, C2], density, G)
\end{lstlisting}
where the inputs of the method are in order the length of the layer ({\ttfamily length}), the complex refractive index of the layer ({\ttfamily n}), the list of the thermal conductivities of each temperature system ({\ttfamily [k1,  k2]}), the list of the specific heats of each temperature system ({\ttfamily [C1, C2]}), the mass density ({\ttfamily density}) and the exchange coupling ({\ttfamily G}).

Thermal conductivities and specific heats can either be numbers or {\ttfamily lambda} functions when they vary with temperature. In case $N=3$, {\ttfamily G} can be a list containing the coupling constants $G_{12}$, $G_{23}$ and $G_{31}$ in this order.

Before running the simulation, a final time must be given by the user through the command
\begin{lstlisting}[language=Python]
# Set final time (in s)
sim.final_time = final_time
\end{lstlisting}

The commands described so far are sufficient to run the simulation. If no further input is given some default values are selected according to the most common conditions or models employed for the experiments.\\
By default the time step is assigned automatically through the procedure illustrated in Sec. \ref{sec: time step}. However the user can define a different time step by 
\begin{lstlisting}[language=Python]
sim.time_step = time_step
\end{lstlisting}
The usual initial condition for all the temperatures $T^i$  by default is $T^i = 300$K. The user can modify the initial condition through the command line
\begin{lstlisting}[language=Python]
# Modify initial condition of i-th temp. (in K)
sim.changeInit( i, T0)
\end{lstlisting}
where {\ttfamily i} varies between 1 and {\ttfamily N}, indicating, that different initial conditions can be set for each subsystem. Furthermore the initial temperature  {\ttfamily T0} can be either a number or a {\ttfamily lambda} function describing the space profile of the temperature $T(x,0)$.

By default the boundary conditions are of modified Neumann type with no heat flux through the boundaries, corresponding to an insulated system. The boundary condition type and value can be modified through the following methods
\begin{lstlisting}[language=Python]
# Change BC type for the i-th temp
sim.changeBC_Type( i, side, Type)
# Change BC value for the i-th temp
sim.changeBC_Value( i, side, BC)
\end{lstlisting}
where again {\ttfamily i} is the index of the temperature system subject to the change, {\ttfamily side} can be either {\ttfamily 'left'} or {\ttfamily 'right'} depending on which side the change is applied to, {\ttfamily Type} is either  {\ttfamily 'dirichlet'} or  {\ttfamily 'neumann'} respectively, and {\ttfamily BC} is a constant or a \texttt{lambda} function providing the value of the boundary condition as the time varies.

Once the input of the data is complete the simulation is executed by the command
\begin{lstlisting}[language=Python]
# Compute temperature map
[x, t, T] = sim.run()
\end{lstlisting}
It yields the {\ttfamily numpy.array} containing the 2+N-dimensional arrays {\ttfamily T} of the temperature of the N subsystems on the space grid  {\ttfamily x} at the times {\ttfamily t}. 

\subsection{Plot Results}

Once the simulation has been executed the results, that is the dynamics of all respective temperatures in space and time, are provided to the user in array form, as described above. However, in order to make the visualization both, easier and quicker for the the user, a separate class with relevant plotting methods has been defined. This gives the user the freedom to output the raw data and do post-processing on their own but also to use the visualization class in order to depict some properties immediately.

\begin{figure}[t]
   \centering
    \includegraphics[scale = 0.4]{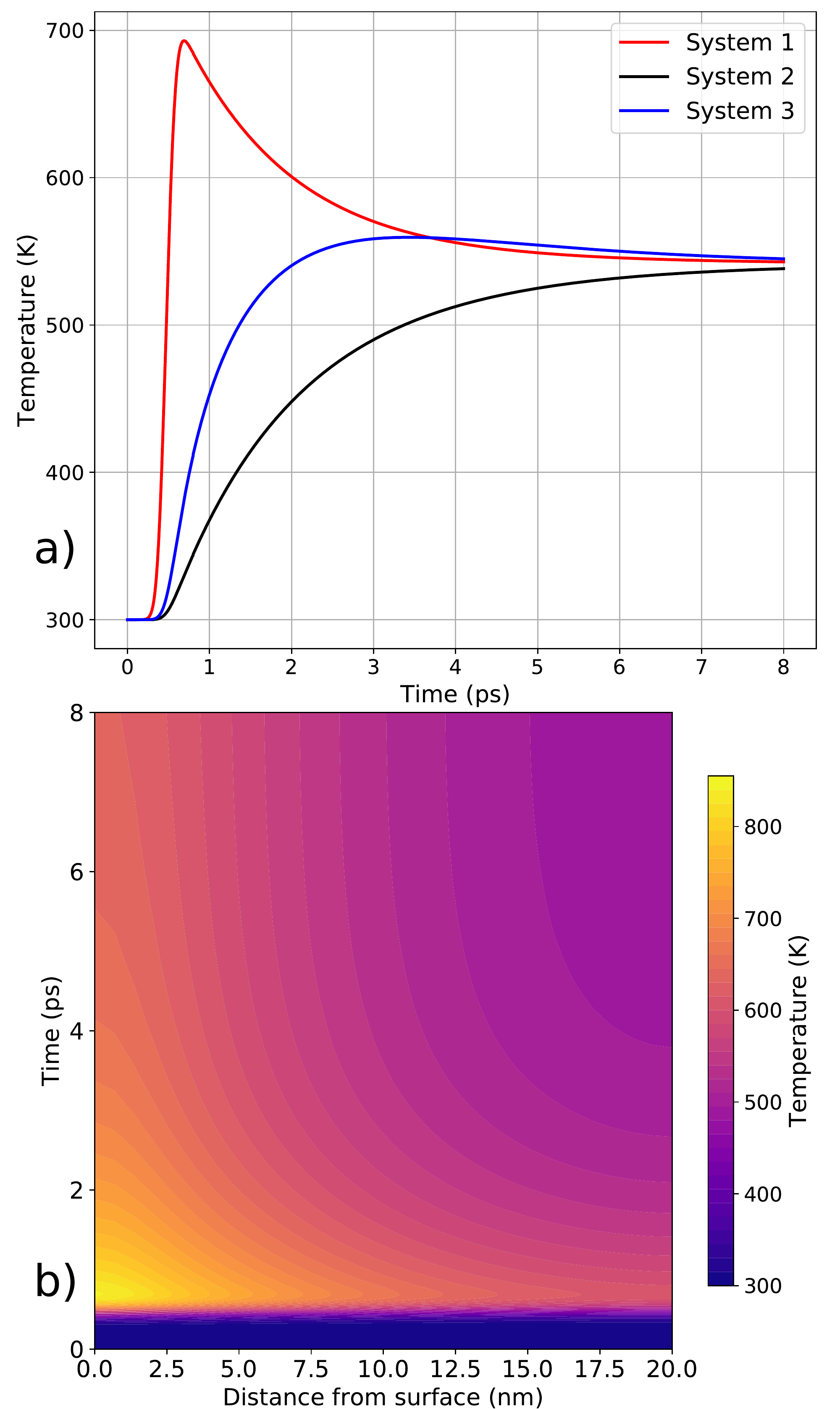}
    \caption{Example of a 3-temperature simulation and the output of two visualization methods: a) {\ttfamily v.average()} and b) {\ttfamily v.contour('1')} described in the text. Here a \SI{20}{nm} ferromagnetic nickel thin film, heated by a femtosecond laser source is under consideration. The physical parameters are taken from Ref. \cite{Bigot}.}
   \label{fig:type4}
\end{figure}

Among others, there are methods to visualize the heating source $S(x,t)$,  a contour plot of the temperature in space and time $T^{E,L,S}(x,t)$, but also some simple ready made post-process routines like the plot of the average temperature of a layer over time, see Fig. \ref{fig:type4}. Moreover the package provides a method to play an animation of how all the temperature systems evolve in time which can be useful either to visually detect the impact of the fundamental mechanisms in the case considered, or for pedagogical use.
Different routines can be called by following the commands
\begin{lstlisting}[language=Python]
# Creating a visual object
v = visual(sim)
# Depicting results and obtaining raw data
source = v.source()
[timegrid,average_temp] = v.average()
v.contour('1')
v.animation(speed,save)
\end{lstlisting}
where the output of the first two visualization routines are arrays, which allow users, to look and process the raw data themselves. The only visualization methods that require input arguments are {\ttfamily v.contour('N')}, 
where {\ttfamily N} can be 1,2 or 3, corresponding to   electron-, the lattice- or the spin- system, and  {\ttfamily animation(speed,save)}, where the speed of the animation can be adjusted and the user can decide whether they want to save the animation, with {\ttfamily save = 1} or not, with {\ttfamily save = 0}.

This makes it easy to quickly check results and analyze properties of the dynamics of the simulation.

\section{Package Validation}\label{sec:Example}

\subsection{Comparison With Other Simulation Tools}\label{sec: Validation&Performance}

In order to validate the results obtained from the NTMpy package, the focus is set on the two main parts of the software. First, the evaluation of the local absorption with respect to the incident laser pulse, and second, the performance of the solver itself. Looking at Eq.~\eqref{eq: N-temperature model}, we are interested of how accurately we are computing the source term $S(x,t)$ responsible for heating and how well our obtained solution $T(x,t)$ matches with commercial software, when experimentally relevant cases are under investigation. 

\begin{figure*}[t]
    \centering
    \includegraphics[width = \textwidth]{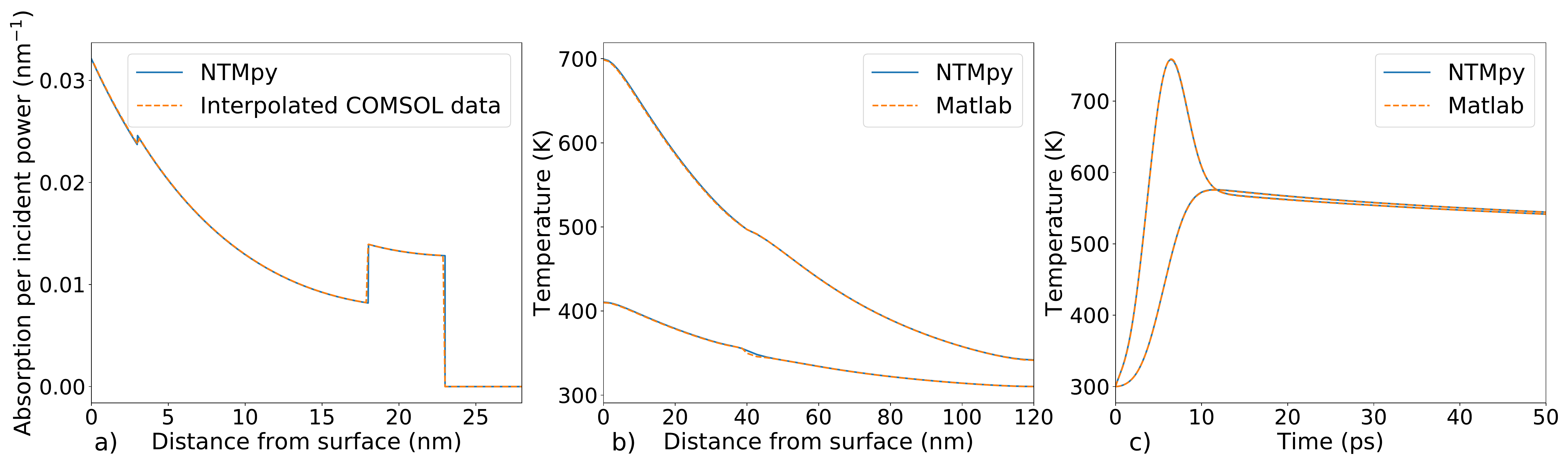}
    \caption{(a) Local absorption per incident laser power computed with NTMpy's TMM-module and with the commercial software COMSOL Multiphysics\texttrademark, Ref. \cite{COMSOL}. For both simulations a three-layer material with identical parameters is considered. Line cuts for two coupled temperature systems $T^{E,L}(x,t)$ at (b) a fixed time $t=t_0$ and at (c) a fixed point in space $x=0$. NTMpy's solutions for both systems (electronic and lattice), computed for a two-layer stack, are compared to  Matlab's built-in partial differential equation {\ttfamily pdepe()} solver, Ref. \cite{MATLAB}.}
    \label{fig: comparison}
\end{figure*}

For the evaluation of the local absorption of $S(x,t)$, we are considering a [Pt 3nm|Co 15nm|Cr 5nm|MgO]- material stack and compute the local absorption per unit incident power, with respect to the distance from the surface, with the implemented transfer matrix method. The obtained profile is compared to the result, obtained from COMSOL Multiphysics\texttrademark, Ref. \cite{COMSOL},  simulations, where the same light and material parameters are considered. In Fig.~\ref{fig: comparison} a), we see, that both, the commercial and our open source software show an overlapping result (mean relative discrepancy $\overline{\delta(x)} = 2.2\%$).

For the evaluation of computation time and accuracy of the obtained solution, we compare $T(x,t)$ to Matlab´s, Ref. \cite{MATLAB}, built in partial differential equation, {\ttfamily pdepe()}, - solver.  Again we consider the same light and material parameters for both software and obtain output that is very much in agreement, with respect to space Fig.~\ref{fig: comparison} b) and time c) dimension.  That is the relative error of the temperature enhancement $\frac{T_{NTMpy}-T_{Matlab}}{T_{Matlab}-300}\leq 6\%$. 
Also the time needed until a solution is obtained from our free package is comparable to Matlab's performance.  

\begin{figure}[h]
\centering
\includegraphics[width=0.9\columnwidth]{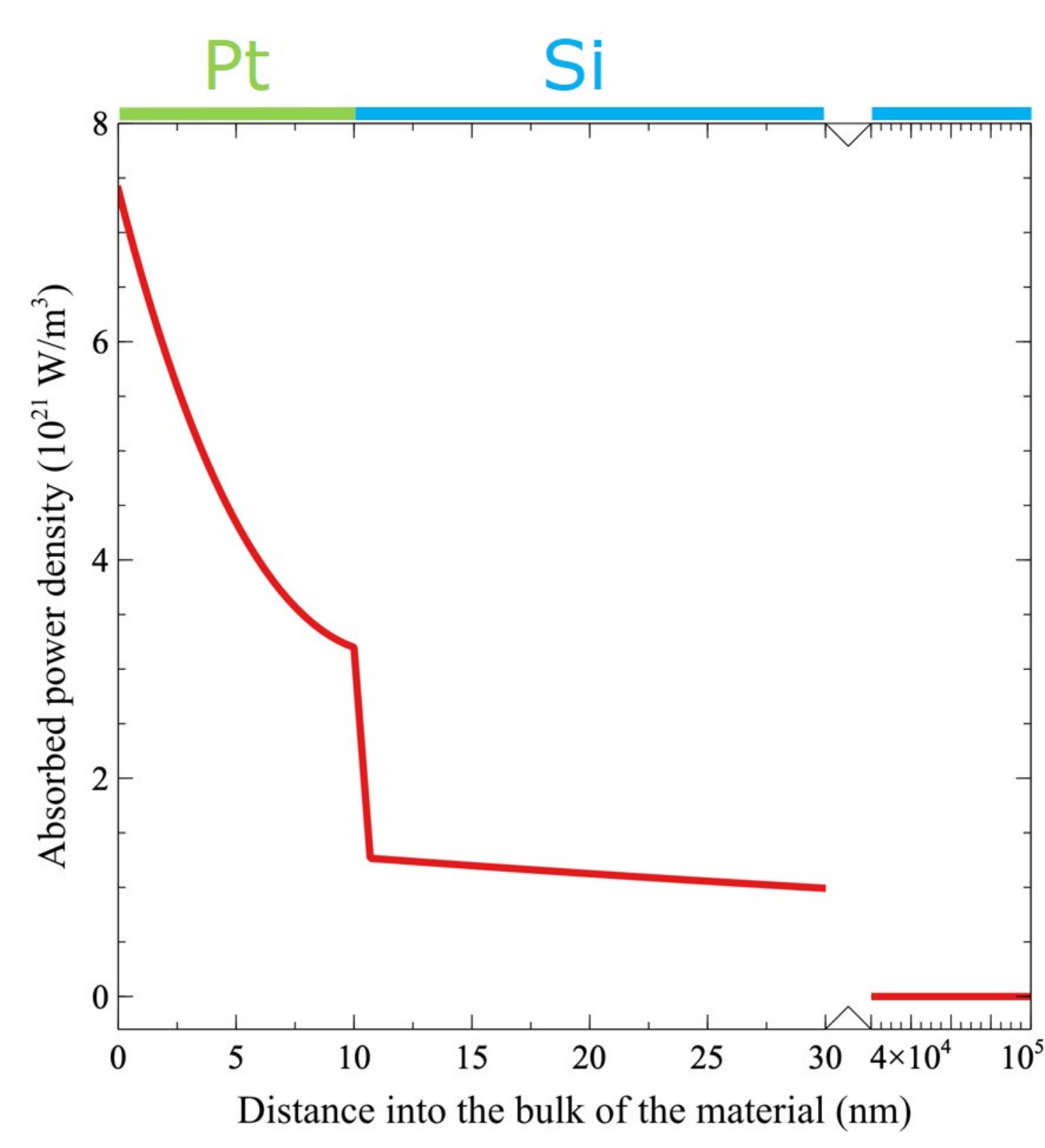}
\caption{Local power absorption for the platinum/silicon system under consideration, calculated via the transfer matrix method implemented in NTMpy. Here a p-polarized laser pulse with a wavelength of \SI{400}{nm}, hitting the sample at a \SI{45}{\degree} angle with a fluence of \SI{6}{mJ/cm^2} and a FWHM of \SI{0.1}{ps} is considered.}
\label{fig: SourceStack}
\end{figure}

\subsection{Comparison With Experiment}

Even though the software is able to solve generic coupled non linear 1-D diffusion equations in the form of Eq. \eqref{eq: N-temperature model}, the focus is laid on heat transfer between systems and transport along the material, within the framework of the two/three temperature model.

If the physical parameters of all the material layers under consideration are known, the software can be used to depict the temperature dynamics in space and time of the selected configuration. This provides an opportunity to be able to detect local phenomena in a material, as opposed to averaged effects, measured in most experiments. However reliable data on certain parameters is sparse and it is a current interest of research to find those properties.

In order to demonstrate the use of this software, we measured the change in reflectivity on a \SI{10}{nm} platinum thin film on a silicon substrate and used the NTMpy software to perform a simulation of the temperature dynamics of the same system. The experimental data was retrieved from a pump-probe transient reflectivity measurement. According to Refs. \cite{HohlfeldTTM, Caffrey}, a linear relationship between the measured change in reflectivity and the change in temperature can be assumed, i.e. $\frac{\Delta R}{R}\propto \frac{\Delta T}{T}$. This makes it possible to compare the experiment to the simulation.

In the simulation, we used the same nominal laser source parameters as in the experiment. That is a \SI{400}{nm}, p-polarized laser pulse with a FWHM of \SI{100}{fs} and a fluence of \SI{6}{mJ/cm^{2}} of which 53\% get absorbed, which is hitting the sample in a 45° angle. The local absorption profile is shown in Fig. \ref{fig: SourceStack}. For the platinum film and the silicon substrate the parameters from Table \ref{tab:2TMParameters} are used.

\begin{table*}[]
    \centering
    \begin{tabular}{|p{0.25\linewidth}|p{0.20\linewidth}|p{0.20\linewidth}|p{0.20\linewidth}|}
    \hline
    Parameter                   & Symbol/ Units&       Platinum      &         Silicon \\ \hline
    Thickness               & l (nm)&10                 & 100000            \\ \hline
    Refractive index (at 400 nm)& $n$ &  1.7176 + i2.844 \cite{RefractivePt}    & 5.5674+0.38612j \cite{RefractiveSi} \\ \hline
    Electron heat conductivity & $k_e (\mathrm{Wm^{-1}K^{-1}}$)& 72 \cite{AmericanInstitute}                  &  130 \cite{SiliconKel}\\ \hline
    Lattice heat conductivity & $k_l (\mathrm{Wm^{-1}K^{-1}}$) & 72 \cite{AmericanInstitute}                    &  $k_l(T)$\cite{Silicon} \\ \hline
    Electron heat capacity  & $C_e(T) (\mathrm{Jkg^{-1}K^{-1}}$) & $740/\rho_{Pt}T_e$ \cite{AmericanInstitute}  & 150/$\rho_{Si}\cdot T_e$ \cite{Silicon}\\ \hline 
    Lattice heat capacity &$C_l (\mathrm{Jkg^{-1}K^{-1}}$) & 2.78E6/$\rho_{Pt}$ \cite{AmericanInstitute}& 1.6E6/$\rho_{Si}$ \cite{Silicon} \\ \hline
    Electron phonon coupling  & $G (\mathrm{Wm^{-3}K^{-1}})$ &2.5E17\cite{Hohlfeld} & 18E17 \cite{CouplingSi}\\ \hline
    \end{tabular}
    \caption{Material parameters used for the TTM simulation in Fig. \ref{fig: PlatinumSilicon_exp_sim}. }
    \label{tab:2TMParameters}
\end{table*}{}

\begin{figure}[h]
\centering
\includegraphics[width=0.9\columnwidth]{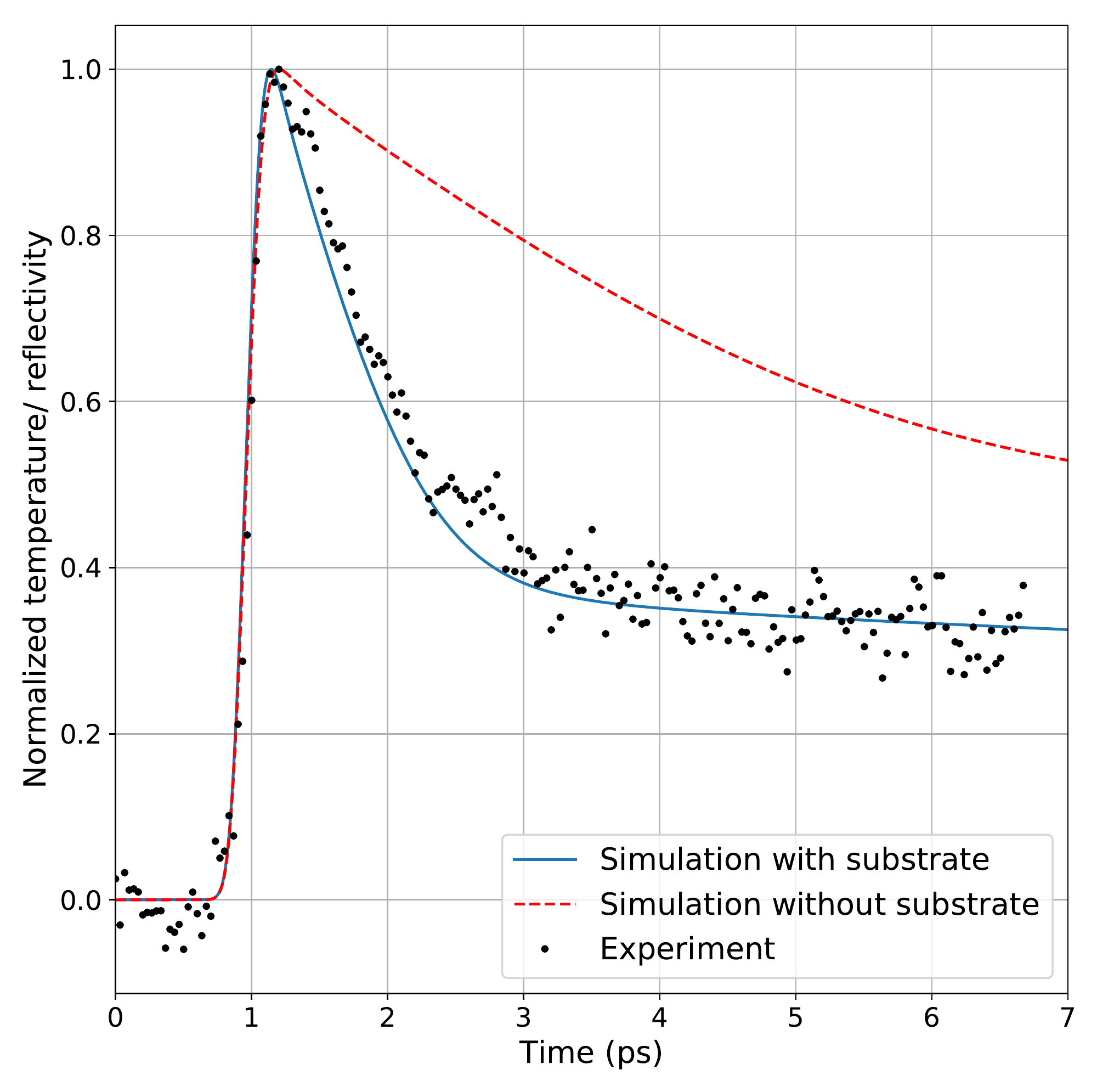}
\caption{Experimental data of a reflectivity measurement (symbols) compared to the simulated temperature dynamics of a platinum thin film on a silicon substrate (solid line) and free-standing (dashed line). A linear relation between the change in temperature and change in reflectivity is assumed, and they are normalized to the maximum change. The simulation parameters for the materials are taken from Table \ref{tab:2TMParameters}.}
\label{fig: PlatinumSilicon_exp_sim}
\end{figure}

The simulation is set up as follows:
\begin{lstlisting}[language=Python]
sim = simulation(2,s)
sim.addLayer(length_Pt,n_Pt,[k_el,k_lat],
                [C_el,C_lat],rho,[G_Pt]) 
sim.addSubstrate("Si")
sim.final_time = 7*u.ps
[x,t,T] = sim.run()
\end{lstlisting}

To compare the simulated data with the experimental case, we computed the exponentially weighted temperature data $T^{E,L}(x,t)$ with respect to the depth of the material. This mimics the  effect of limited optical penetration depth of the probe laser. In Fig.~\ref{fig: PlatinumSilicon_exp_sim}, we show a comparison between a simulation with and without the silicon substrate.

The results, depicted in Fig.~\ref{fig: PlatinumSilicon_exp_sim} show that the multilayer solution obtained from the NTMpy software is able to reproduce the experimental data to a very good degree. Furthermore, one can clearly see that disregarding the substrate in the simulation, which is the simplest way to solve the two-temperature model, leads to a considerably different results. In this case, a four times larger electron-lattice coupling constant from $G = 2.5\cdot10^{17}(\mathrm{Wm^{-3}K^{-1}})$ to $G \approx 11\cdot10^{17}  (\mathrm{Wm^{-3}K^{-1}})$ would be needed to reestablish agreement with the experiment.

\section{Conclusion}\label{sec:conclusion}

We implemented NTMpy, an open source Python based software package for solving coupled parabolic differential equations in one dimension. Along with the mathematical background of the algorithm, we introduced the structure and the work flow of the program. The object oriented way in which the program is designed gives the user the freedom to run tailor-made simulations, without having to worry about coding details, which are automatized or ready-made. This helps to focus more on the output and analysis of the fundamental dynamics. Together with the visualization class, the software can not only be used by researchers to investigate relaxation and diffusion dynamics, but also for educational purpose. Finally, a comparison of the output of NTMpy with the one of both commercial software and also of new experimental data, demonstrated the numerical reliability of the software, and its ability to produce physically realistic results.

\section*{Acknowledgements}
L.A. acknowledges support from the Swedish Research Council (VR), Grant: 2018-04611. V.U. and S.B. acknowledge support from the European Research Council, Starting Grant 715452 ``MAGNETIC-SPEED-LIMIT''.

\bibliographystyle{elsarticle-num}

\end{document}